\documentclass[twocolumn,,amsmath,amssymb]{revtex4-1}

\pdfoutput=1

\usepackage{graphicx}% Include figure files
\usepackage{bm}% bold math
\newcommand{\be}{\begin{equation}}
\newcommand{\ee}{\end{equation}}
\newcommand{\bfig}{\begin{figure}}
\newcommand{\efig}{\end{figure}}

\usepackage{graphicx}
\usepackage{textgreek}% For use of "\textOmega, etc.": Greek letters in normal mode.
\usepackage{lineno}
\usepackage{lipsum}
\usepackage{xcolor}

\begin{document}

\title{Singular angular magnetoresistance and sharp resonant features in a high-mobility metal with open orbits, ReO$_3$}

\author{Nicholas P. Quirk$^1$, Loi T. Nguyen$^2$, Jiayi Hu$^1$, R. J. Cava$^2$, N. P. Ong$^1$}
\affiliation{Department of Physics$^1$ and Department of Chemistry$^2$, Princeton University, Princeton, NJ 08544
}

\date{\today}

\begin{abstract}
We report high-resolution angular magnetoresistance (AMR) experiments performed on crystals of ReO$_3$ with high mobility (90,000 cm$^2$/Vs at 2 K) and extremely low residual resistivity (5-8 n$\Omega$cm).  The Fermi surface, comprised of intersecting cylinders, supports open orbits. The resistivity $\rho_{xx}$ in a magnetic field $B$ = 9 T displays a singular pattern of behavior. With $\bf E\parallel \hat{x}$ and $\bf B$ initially $\parallel\bf\hat{z}$, tilting $\bf B$ in the longitudinal $k_z$-$k_x$ plane leads to a steep decrease in $\rho_{xx}$ by a factor of 40. However, if $\bf B$ is tilted in the transverse $k_y$-$k_z$ plane, $\rho_{xx}$ increases steeply by a factor of 8. Using the Shockley tube integral approach, we show that, in ReO$_3$, the singular behavior results from the rapid conversion of closed to open orbits, resulting in opposite signs for AMR in orthogonal planes. The floor values of $\rho_{xx}$ in both AMR scans are identified with specific sets of open and closed orbits. Also, the ``completion angle'' $\gamma_c$ detected in the AMR is shown to be an intrinsic geometric feature that provides a new way to measure the Fermi radius $k_F$. However, additional sharp resonant features which appear at very small tilt angles in the longitudinal AMR scans are not explained by the tube integral approach.
\end{abstract}

\maketitle

%\linenumbers 

\section{Introduction}
The past decade has witnessed renewed interest in semi-metals and metals that exhibit unusually high carrier mobilities. In the Dirac semimetal Cd$_3$As$_2$, the mobility $\mu$ can attain 10$^7$ cm$^2$/Vs~\cite{Liang}. The large-$\mu$ semimetal WTe$_2$ displays non-saturating magnetoresistance in magnetic fields up to 60 T~\cite{MAli}. The Weyl semimetals TaAs, NbAs and NbP have mobilities exceeding 150,000 cm$^2$/Vs. These enhanced $\mu$ may result from a very small effective mass in the vicinity of avoided band crossings and protection from carrier scattering. In metals, the Fermi energy is remote from such band crossings, but high-mobility candidates have also been identified, for e.g. PdCoO$_2$, PtCoO$_2$~\cite{NNandi,Coldea} and Pd$_3$Pb~\cite{Mitchell}. For Fermi surfaces that are mutliply connected, angular magnetoresistance (AMR) is a powerful tool for unravelling how connectivity affects transport. Although AMR is most frequently employed to map the angular variation of the Shubnikov de Haas (SdH) period, for e.g., in Sr$_2$RuO$_4$~\cite{Bergemann} and the Bechgaard salts, it can also uncover surprising features unrelated to SdH oscillations. The Yamaji angle detected in the Bechgaard salts is a well-known example~\cite{Yamaji,Kagoshima}. A more recent example is the existence of ultra-narrow peaks in the AMR of the magnetic Weyl semimetal CeAlGe when $\bf B$ is aligned with symmetry axes~\cite{Checkelsky}.

Here we report novel features observed in the AMR of crystals of ReO$_3$ that exhibit extremely low residual resistivities. ReO$_3$ is the archetypal example of a metal in which the Fermi surface (FS) forms a three-dimensional (3D) jungle-gym network of intersecting cylinders plus two small closed surfaces~\cite{Matthiess,Cora,Stachiotti}. Early experiments on ReO$_3$ are reported in Refs. \cite{Marcus,Graebner,RPhillips,Pearsall}. From a modern viewpoint, ReO$_3$ has some interesting features. The lattice structure, comprised of a Re ion surrounded by six nearest neighbor O ions, is the simplest expression of a 3D Lieb lattice~\cite{Lieb}. A hallmark of Lieb lattices is the existence of flat bands caused by wave-function interference~\cite{Wu,Regnault}. In ReO$_3$, flat bands are prominent along $X$-$M$, but they lie too far from the Fermi level (by 1 eV) to affect transport directly.

We have grown crystals in which the residual resistivity $\rho_{00}$ is 5 to 8 n$\Omega$cm at 2 K (6-10 times lower than in ultra-pure Au). At 2 K, $\mu$ is estimated to be 90,000 cm$^2$/Vs. In these crystals, we have uncovered a singular feature in the AMR. With axes $x$, $y$ and $z$ fixed parallel to the cylinders' axes, and the electric field $\bf E$ $\parallel \bf\hat{x}$ (Fig. \ref{fig1}), we observe the longitudinal resistivity $\rho_{xx}$ to decrease by a factor of $\sim$40 when $\bf B$ (fixed at 9 T) is tilted towards $\bf E$. However, if $\bf B$ is tilted in the plane orthogonal to $\bf E$, $\rho_{xx}$ exhibits a 10-fold increase. The extreme anisotropy in the response of $\rho$ to slight angular deviations from the singular point $(\theta,\chi) = (0,0)$ ($\bf B\parallel \hat{z}$) has not been reported previously in any metal to our knowledge. All the AMR curves investigated (as well as the Hall response) display a sharp discontinuity at a characteristic angle $\gamma_c\simeq \,29^\circ$.
Moreover, we observe weak features in the scans vs. $\theta$ (sharp resonances) suggestive of enhanced scattering at specific tilt angles 1.1$^\circ$ and 2.2$^\circ$.

We describe a semiclassical model based on open orbits on the jungle-gym Fermi surface (FS) that emphasizes the connectivity of the orbits in tilted $\bf B$ and the key role of orbital links that convert closed to open orbits. The model accounts for the opposite signs of the AMR vs. $\theta$ and $\chi$, as well as the physical meaning of $\gamma_c$ which we call the ``completion'' angle. However, it is inadequate for explaining the cusp-like sensitivity at very small tilt angles or the appearance of sharp resonances. 

\section{Experimental Results}
\begin{figure*}[h]
	\includegraphics[width=0.9\textwidth]{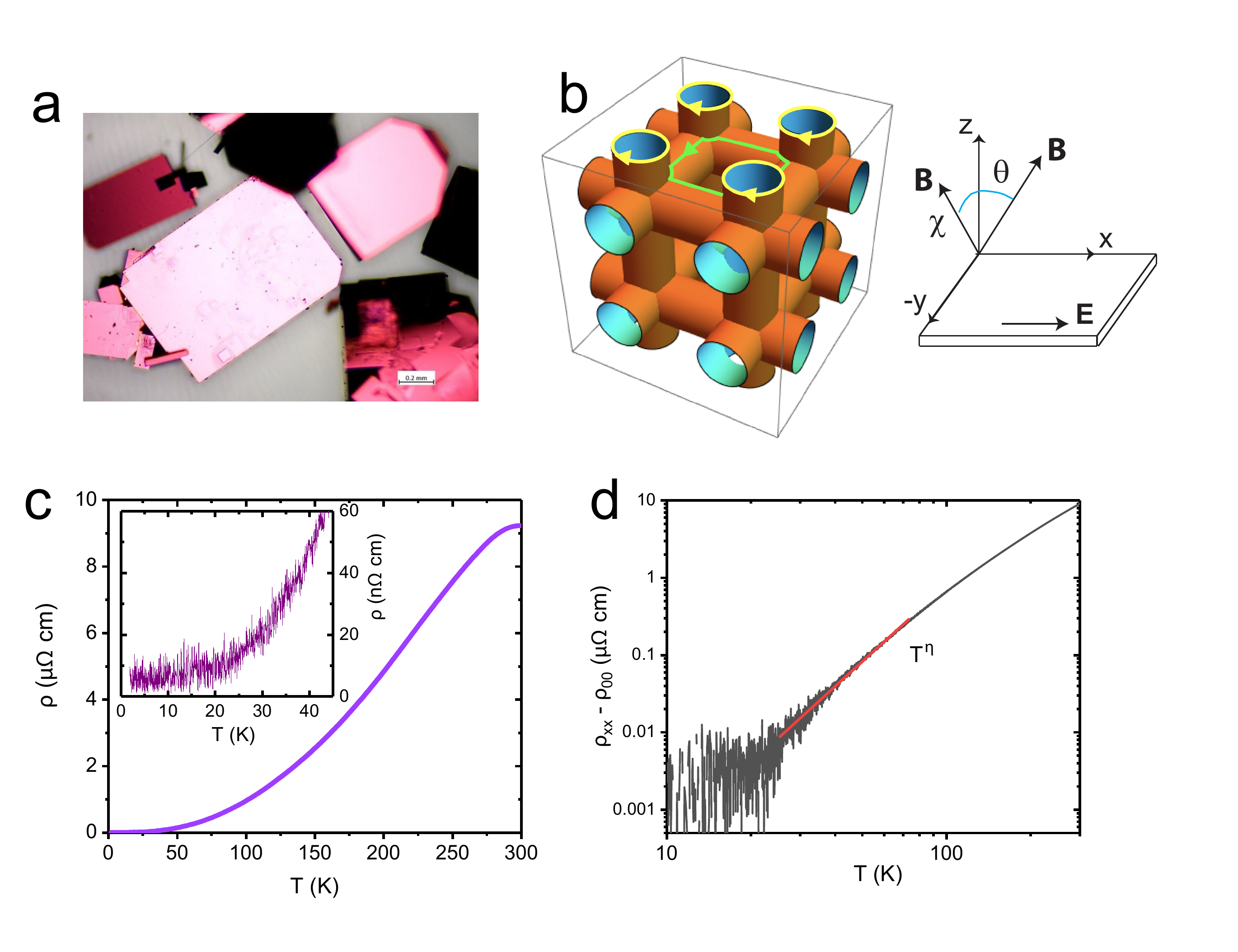}
	\caption{(a) Crystals of ReO$_3$ showing characteristic brilliant pink hue in reflected light. The cubic cell parameter is 3.748 \AA. (b) Sketch of the jungle-gym FS with 4 electron-like closed cyclotron orbits (yellow loops) and a hole-like closed orbit (green). The inset shows the field tilt-angles $\theta$ and $\chi$ relative to axes ($x,y,z$). (c) Plot of the resistivity $\rho$ vs. $T$ with $B = 0$. The residual value $\rho_{00}$, measured in 4 crystals, is 5-8 n$\Omega$cm (inset). (d) Log-log plot of $\Delta\rho$ vs. $T$ where $\Delta\rho(T) = \rho(T) - \rho_{00}$. A linear fit (red line) over $20 < T< 80$ K gives $\Delta\rho = T^\eta$ with $\eta =3.1\pm 0.2$.}
	\label{fig1}
\end{figure*}

Crystals of ReO$_3$ were grown by double-pass chemical vapor transport. A silica tube of inner diameter 14 mm and length 30 cm was loaded with 1 g of ReO$_3$ powder and 25 mg of iodine flakes and sealed under vacuum. The tube was inserted into a 3-zone horizontal tube furnace in which the temperature was slowly raised over 6 h to 500$^\circ$C (hot end) and 450$^\circ$C (cool end). After 4 days of vapor transport, the furnace was cooled over 10 h to 290 K. Vapor transport was then repeated to enhance the crystal purity. Large red plate-like crystals up to 1 cm on a side were harvested at the cold end (Fig. \ref{fig1}a). The phase purity and crystal structure of ground crystals were determined by powder x-ray diffraction using a Bruker D8 Advance Eco with Cu Kα radiation and a LynxEye-XE detector.

Figure \ref{fig1}b shows a sketch of the jungle gym FS, using the value of the Fermi radius $k_F$ from Ref.~\cite{RPhillips}. In the profile of the zero-$B$ resistivity $\rho$ vs. $T$ (Fig. \ref{fig1}c), $\rho$ maintains its ultralow residual value $\rho_{00}$ (inset) to an unusually high $T\sim 20$ K, implying that phonon scattering is suppressed until $T$ exceeds $\sim 20$ K. The residual resistivity ratio $\rho(300 \,\rm{K})/\rho_{00}$ is 1,500. The $T$-dependent part $\Delta\rho(T) = \rho(T)-\rho_{00}$ fits well to $T^\eta$ up to 80 K (Fig. \ref{fig1}d) with an exponent $\eta \simeq 3.1\pm 0.2$ much reduced from that in the Bloch law ($T^3$ vs. $T^5$). 

We selected crystals with optimal rectangular shape (1.0 $\times$ 0.5 mm$^2$ in area) and mechanically polished the broad faces with fine sandpaper to reduce the thicknesses to 80-100 um. The edges of the broad face are aligned (to a precision of $\pm 1^\circ$) with $k_x$ and $k_y$ of the lattice. In all field-tilt measurements, we define the $x$, $y$, and $z$ axes to be anchored to the $k_x$, $k_y$ and $k_z$ axes of the lattice, respectively (Fig. \ref{fig1}b). Both the electric field $\bf E$ and the (spatially averaged) current density $\bf \langle J\rangle$ are $\parallel\bf \hat{x}$. The contact resistances of the Ag paint contacts were under 2 $\Omega$.

The sample platform was tilted using a horizontal rotator in a Quantum Design PPMS equipped with a 9-Tesla magnet. The field tilt angles, $\theta$ and $\chi$ defined in Fig. \ref{fig1}b were measured with a transverse Hall sensor (Lakeshore HGT 2101-10) to a resolution of $\pm 0.03^\circ$. The 4-probe measurements of resistances were performed using a Keithley 6221 DC current source and 2182a nanovoltmeter in Delta mode using current pulses of 5-10 mA. 

\begin{figure*}[h]
	\includegraphics[width=0.9\textwidth]{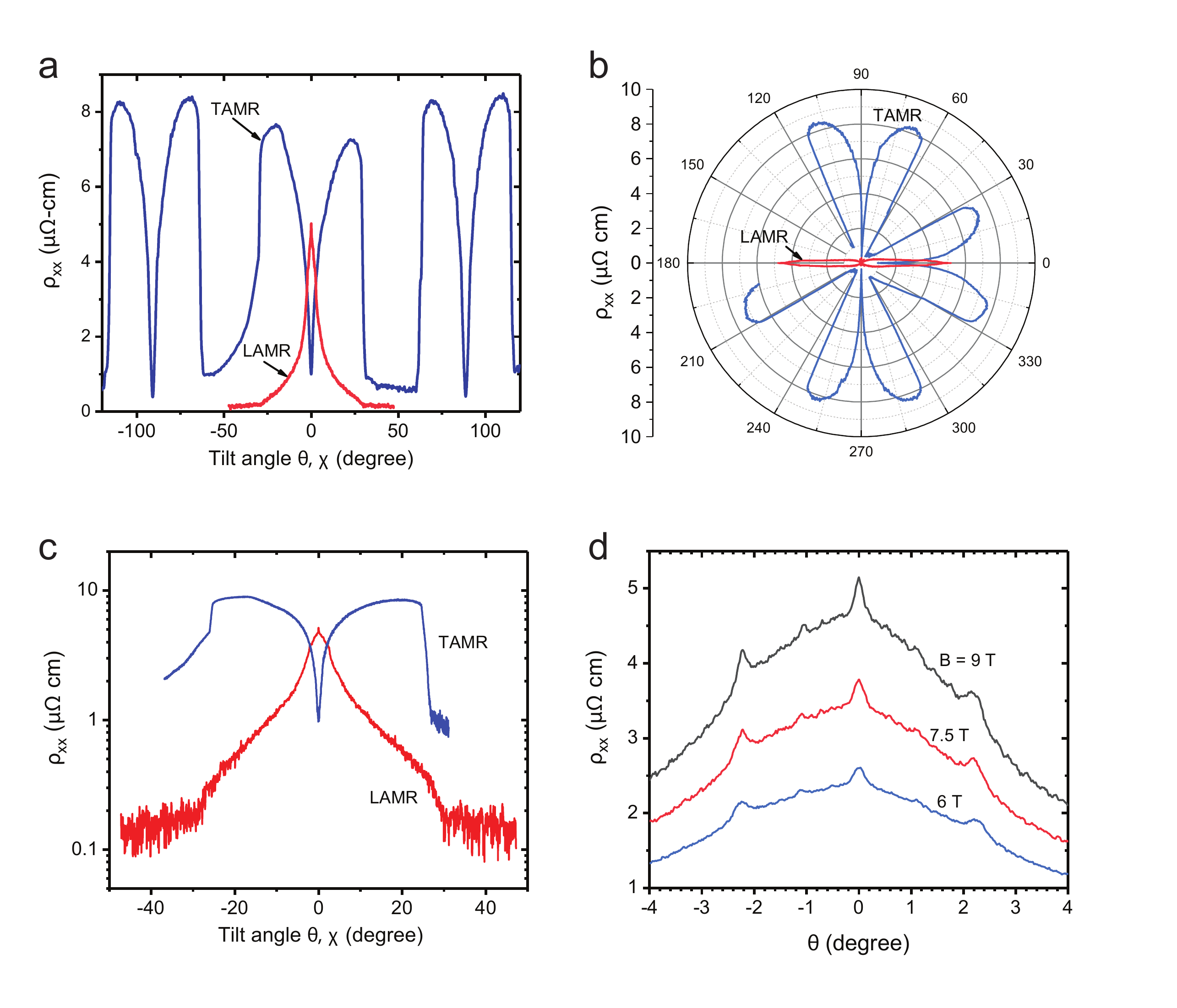}
	\caption{Panel (a): The singular, anisotropic angular magnetoresistance $\rho_{xx}(\theta,\chi)$ measured at $T$ = 1.9 K with $\bf E\parallel \hat{x}$ and $\bf |B|$ fixed at 9 T. The LAMR curve (in red) plots $\rho_{xx}(\theta,0)$ vs. $\theta$ with $\bf B$ lying in the (longitudinal) $x$-$z$ plane at angle $\theta = \angle({\bf B,\;\hat{z}}$). The TAMR curve (blue) plots $\rho(0,\chi)$ vs. $\chi$ with $\bf B$ in the transverse $y$-$z$ plane at angle $\chi = \angle({\bf B,\;\hat{z}})$. 
A slight misalignment causes a weak breaking of mirror symmetry about $\chi = 0$ or $\theta = 0$ (see text). The singular AMR complicates determination of $\rho_{xx}(\theta,\chi)$ at $(\theta, \chi) = (0,0)$. Panel (b) shows the polar plot of the TAMR and LAMR curves. The TAMR curve (blue) displays $C_4$ symmetry. However, the LAMR curve (red) exhibits $C_2$ symmetry because, with $\bf E$ fixed $\parallel\bf\hat{x}$, $\rho_{xx}(0,0)\gg \rho_{xx}(\pi/2,0)$ (the latter is equal to $\rho_{zz}(0,0)$).  
Panel (c) is an expanded view of the curves of LAMR (red) and TAMR (blue) in semi-log plot. The TAMR curve shows a steep decrease at the completion angle $\gamma_c$. The step decrease in the LAMR curve is milder but still well resolved. 
Panel (a): Expanded view of the LAMR curve $\rho_{xx}(\theta,0)$ at 1.9 K with $\bf |B|$ fixed at 6 T (blue curve), 7.5 T (red) and 9 T (grey). In all three curves, sharp resonant features are observed at $\theta$ = 0, $\pm 1.1^\circ$ and $\pm 2.2^\circ$. 
}
	\label{fig2}
\end{figure*}

When $\bf B$ is tilted by $\theta$ in the longitudinal $x$-$z$ plane with $\chi$ fixed at 0, $\rho_{xx}(\theta, 0)$ displays sharp maxima at $\theta = 0$ and $180^\circ$. Figure \ref{fig2}a plots $\rho_{xx}(\theta,0)$ vs. $\theta$ measured at $T$ = 1.9 K (red curve). We call this the longitudinal AMR (LAMR) curve. In the polar plot, the LAMR curve describes two very narrow plumes directed along $\theta = 0$ and 180$^\circ$ (red curves in Fig. \ref{fig2}b). An expanded view of the LAMR curve is shown in semilog scale in Fig. \ref{fig2}c. As $\theta$ increases from 0, $\rho_{xx}$ decreases steeply by a factor of $\sim 40$ (semilog plot in Fig. \ref{fig2}c). A characteristic angle $\gamma_c\sim 29^\circ$ (which we call the ``completion'' angle) is prominently seen in all AMR curves investigated. In the LAMR scan, $\rho_{xx}(\theta,0)$ displays a rounded step-drop to the ``floor'' value $\rho^{L,fl}$, where it remains until $\theta\to 150^\circ$. We have $\rho^{L,fl} \simeq 20\times \rho_{00}$.

The transverse AMR (TAMR) curve plotting $\rho_{xx}(0,\chi)$ vs. $\chi$ with $\bf B$ lying in the transverse $y$-$z$ plane, are radically different (blue curve in Fig. \ref{fig2}a). At small tilt angle ($|\chi|<15^\circ$), $\rho_{xx}$ increases steeply to a peak value 8-10$\times$ higher than at $\chi=0$. Further increase of $\chi$ to $\gamma_c$ leads to a steep decrease to a resistivity floor value $\rho^{T,fl}$ that is $10\times$ larger than the floor value $\rho^{L,fl}$ in the LAMR (see the semilog plot in Fig. \ref{fig2}c). We estimate $\rho^{T,fl}= 4.5\times\rho^{L,fl} \gg \rho_{00}$.
The polar plot of the TAMR curve (blue curve in Fig. \ref{fig2}b) shows an 8-petal floral pattern with $C_4$ symmetry weakly broken by misalignment.

In principle, the sharp maximum in $\rho_{xx}$ at $\theta = 0$ in the LAMR curve must equal the minimum in the TAMR at $\chi = 0$. In our experiment, however, a residual misalignment leads to a difference of a factor of 4. The singular behavior in the vicinity of $(\theta, \chi) = (0,0)$ amplifies errors caused by angular misalignments of $\pm 1^\circ$ (the difficulty is roughly similar to aligning the tips of two sharp needles). The traces in Fig. \ref{fig2} result from progressive alignment improvements in repeated scans. The misalignment also accounts for slight deviations from $C_4$ symmetry in the polar plot of the TAMR curve.

Returning to the LAMR curve, we resolve small, very narrow, resonant features at small $\theta$. The expanded view in Fig. \ref{fig2}d displays three LAMR scans measured at 1.9 K with $\bf|B|$ fixed at 6, 7.5 and 9 T. In each curve, $\rho_{xx}$ displays distinct peaks with ultra-narrow widths ($\sim 0.1^\circ$) centered at $\theta = 0$, $\pm 1.1^\circ$ and $\pm 2.2^\circ$. The peak amplitudes are strongest at 0$^\circ$ and $\pm 2.2^\circ$. Because their angular positions are independent of $B$, they are unrelated to quantization of the magnetic flux. We discuss their origin below.

\begin{figure*}[h]
	\includegraphics[width=0.5\textwidth]{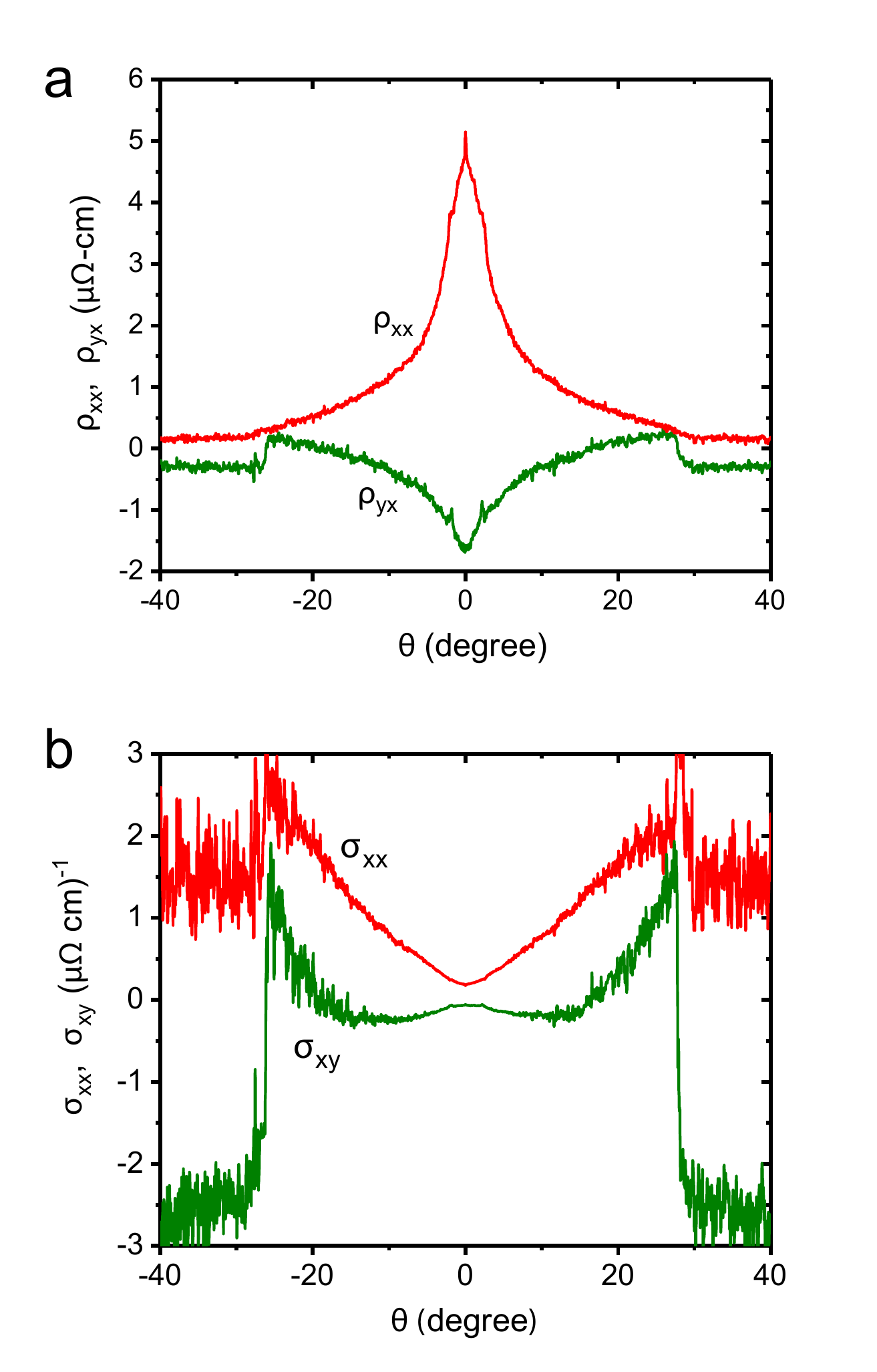}
	\caption{ Panel (a): Comparison of the angular Hall resistivity $\rho_{yx}(\theta,0)$ (green curve) and $\rho_{xx}(\theta,0)$ (red curve) measured vs. $\theta$ (setting $\chi = 0$) at 1.9 K with $\bf|B|$ fixed at 9 T. Initially, $\rho_{yx}$ is electron-type at $\theta = 0$, but changes to hole-like near $16^\circ$. At $\gamma_c$, $\rho_{yx}$ undergoes a step-wise change, involving a second sign-change. The curves for the inferred conductivity $\sigma_{xx}$ (red curve) and Hall conductivity $\sigma_{xy}$ (green) are plotted in Panel (b). At small $\theta$, $\sigma_{xy}$ is negative. Near 16$^\circ$, it changes sign and increases steeply before suffering a large discontinous jump at $\gamma_c$ to return to negative values.
}
	\label{fig3}
\end{figure*}

To complement the longitudinal resistivity, we have also performed Hall measurements.
In Fig. \ref{fig3}a, the green curve plots the angular Hall resistivity $\rho_{yx}(\theta,0)$ vs. $\theta$ in the LAMR experiment ($\rho_{yx}$ depends on $B\cos\theta$ so it is even in $\theta$).  At the angle $\gamma_c$, $\rho_{yx}$ displays a remarkable step-decrease that involves a sign change. Inverting the resistivity matrix $\rho_{ij}(\theta,0)$, we obtain the conductivity matrix $\sigma_{ij}(\theta,0)$. The curves of $\sigma_{xx}$ (red) and $\sigma_{xy}$ (green) are plotted in Fig. \ref{fig3}b. As $\theta$ increases from 0, the conductivity $\sigma_{xx}(\theta,0)$ increases monotonically up to $\gamma_c$, above which it becomes nearly independent of $\theta$. The more interesting Hall curve $\sigma_{xy}(\theta,0)$ is initially negative at $\theta = 0$. It displays a broad minimum near 12$^\circ$ and then increases steeply to positive values above 16$^\circ$. At $\gamma_c$, however, $\sigma_{xy}$ suffers a giant discontinuity, ending back at a large negative value that slowly increases in magnitude as $\theta\to 45^\circ$.

In our analysis (next section), we have focused on understanding the diagonal conductivity element $\sigma_{xx}$. The Hall conductivity $\sigma_{xy}$ is more difficult to analyze because the competing hole-like and electron-like contributions demand better estimates of the Hall currents. The interesting Hall behavior is deferred for further investigation.

\section{Semiclassical model}
Given the ${\cal C}_4$ symmetry of the lattice, the sign difference of the AMR scans vs. $\theta$ and $\chi$ and their steep variations are unexpected at first glance. We show that the Shockley tube-integral approach~\cite{Ziman} can account qualitatively for the sign difference and floor values observed. Although AMR curves are usually difficult to calculate, there are several mitigating factors in this material. \emph{Ab initio} calculations~\cite{Matthiess,Cora,Stachiotti} reveal that the cylinders have uniform cross-sections which simplifies the evaluation of the tube integral. Moreover, the condition $\mu B\gg 1$ ensures that the cylinders dominate the conductivity matrix element $\sigma_{xx}$. (As discussed later, the sharp ``resonant'' features appearing in LAMR seem to require a more sophisticated treatment.)

In a magnetic field, $\sigma_{ab}$ is given by the Shockley tube integral (see Appendix) 
\begin{eqnarray}
\sigma_{ab} = \frac{2e^2}{(2\pi)^3\hbar^2} \int \frac{m^*}{\omega_c} {\cal C}_{ab}\; dk_H,
%\sigma_{ab} &=& \frac{2e^2}{(2\pi)^3\hbar^2} \int dk_H\,\frac{m^*\tau}{2\pi}
%\oint d\phi\oint d\phi' \,\times \nonumber\\ 
   %         && \left[v_a(\phi)v_b(\phi-\phi') e^{-\alpha\phi'} \right]. 
\label{eq:sigma}
\end{eqnarray}
with the velocity-velocity correlator ${\cal C}_{ab}$ given by
\begin{eqnarray}
{\cal C}_{ab} &=& \left(\frac{\hbar k_F}{m_0}\right)^2\frac{1}{(1-e^{-2\pi\alpha})} \times\nonumber\\
  & & \int_0^{2\pi} d\phi \int_0^{2\pi}d\phi' \; v_a(\phi)v_b(\phi-\phi')\; e^{-\alpha\phi'}.
\label{eq:Cab}
\end{eqnarray}
where $\bf v(k)$ is the group velocity and $\alpha = (\omega_c\tau)^{-1}$.

We approximate the FS as three intersecting cylinders (radius $k_F$), $C_x$, $C_y$ and $C_z$, with axes along $\bf\hat{x}$, $\bf\hat{y}$ and $\bf\hat{z}$, respectively (Fig. \ref{fig4}a). 

We assume $\bf E\parallel \hat{x}$ throughout. It is convenient to denote the conductivity of an isolated cylinder in zero $B$ as
\be
\sigma^{(1)}_0 = n^{(1)} e\mu,
\label{sigma1}
\ee
where $n^{(1)}$, the carrier density enclosed within the cylinder, is given by
\be
n^{(1)} = 2\frac{\pi k_F^2}{(2\pi)^3}(K-2k_F),
\label{n1}
\ee
with $K = 2\pi/a$ and $a$ is the primitive lattice spacing. In a tilted $\bf B$, Eq. \ref{scyl} in the Appendix gives for $C_y$ (in isolation) the conductivity $\sigma^{Cy}_{xx} = \sigma^{(1)}_0/(1+(\mu B_{y})^2)$.

Including both $C_y$ and $C_z$, the measured residual resistivity at $B=0$ is then 
$1/\rho_{00} = 2n^{(1)}e\mu$. With $K \simeq 4k_F$, we find $n^{(1)} \simeq 0.75\times 10^{22} \;\rm{cm}^{-3}$, which yields $\mu = 90,000$ cm$^2$/Vs.  This estimate is in accord with the Hall resistivity $\rho_{yx}$, which exhibits a peak at $B$ = 0.08 T at 2 K with $\bf B\parallel\hat{z}$.

We next consider open orbits. In a tilted $\bf B$, a wave packet on the FS moves along an orbit (red curves in Fig. \ref{fig4}a) defined by the intersection of a plane normal to $\bf B$ (pale blue plane) and the FS. As drawn, the right-moving wave packet on cylinder $C_y$, loops under $C_x$ (dashed curve) before resuming its straight-line path on $C_y$, whereas the left-moving wave packet in the companion orbit loops over $C_x$. In the high-field limit, such open orbits, with non-vanishing $v_x$, dominate the conductivity $\sigma_{xx}$.

With $\bf B$ strictly $\parallel \bf\hat{z}$, the orbits on the cylinder $C_z$ are closed and electron-like. The orbits on cylinders $C_x$ and $C_y$ are also closed (apart from a negligible subset at the top and bottom of $C_x$ and $C_y$ for which $v_x = 0$). However, they are hole-like (comprised of alternating straight segments on $C_x$ and $C_y$). Because of the high mobility, the contributions of the closed hole orbits on cylinders $C_x$ and $C_y$ to $\sigma_{xx}$ decrease as $1/B^2$ when $\mu B\gg 1$. The absence of open orbits causes the resistivity to increase monotonically in the large-$B$ regime, as observed. Our analysis focuses on the conversion of closed to open orbits for states on $C_x$ and $C_y$. The cylinder $C_z$ is less important for the AMR. However, it plays the dominant role in the angular Hall conductivity $\sigma_{xy}(\theta,0)$ (Fig. \ref{fig3}b), which we leave for a future study.

\subsection{LAMR}
In the LAMR experiment, we observe a dramatic increase in $\sigma_{xx}$ when $\bf B$ is tilted even slightly in the longitudinal $k_x$-$k_z$ plane. To show that this results from a sharp increase in the fraction of open-orbit states, we consider the set of planes normal to $\bf B$. Figure \ref{fig4}b shows cross-sections of three $C_y$ cylinders separated by $K=2\pi/a$ in the repeated zone scheme, together with two planes at the tilt angle $\theta$. The planes that are tangential to the outer cylinders (blue lines) intersect the middle cylinder to define two FS arcs hosting open-orbit states (thick green arcs in Fig. \ref{fig4}b). A wavepacket prepared initially on the left green arc on $C_y$ loops under $C_x$ (as a ``looped segment'') then alternates between straight-line segments on $C_y$ and looped segments on $C_x$ (thick red curves in Fig. \ref{fig4}a). Conversely, if the initial state lies outside the green arcs, the wavepacket runs into a neighboring $C_y$ before it can complete a loop on $C_x$. These states, lying in the ``shadow'' cast by adjacent cylinders, remain trapped in closed hole-like orbits.

The looped segments on $C_x$ are crucial for linking straight segments on $C_y$ into open orbits even though they themselves do not contribute to $\sigma_{xx}$. Increasing $\theta$ converts more of the states on $C_x$ to looped segments (as the fraction in the shadow decreases). This results in a sharp increase in the fraction of states on $C_y$ that become open orbits. Hence $\sigma_{xx}$ increases rapidly with $\theta$.

\subsection{Completion Angle}
The increase in $\sigma_{xx}$ ends abruptly when the blue line becomes the inner tangent to adjacent cylinders (red dashed line in Fig. \ref{fig4}b) at the ``completion angle'' $\gamma_c$ given by 
\be
\sin\gamma_c = \frac{2k_F}{K}.
\label{eq:gammac}
\ee
The completion angle provides a direct way to measure $k_F$.

As mentioned, $\rho_{xx}$ abruptly drops to its ``floor'' value at $\gamma_c\sim 29^\circ$ and stays there until $\theta$ exceeds $150^\circ$ (Fig. \ref{fig2}c). Using Eq. \ref{eq:gammac}, we find that $k_F/K$ = 0.25 in good agreement with an earlier de-Haas-van Alphen experiment~\cite{RPhillips} which reported $k_F/K$ =0.23.
The negative LAMR profile provides a new way to measure $k_F$ in ReO$_3$. In both the Hall scan and the TAMR experiment, the step-changes at $\gamma_c$ are much more pronounced.

In the floor interval $\gamma_c<\theta < \pi-\gamma_c$, nearly all the states on $C_y$ belong to open orbits (the green arcs extend over the entire cross section), so their contributions revert to their zero-$B$ values. The states on $C_z$ are all in closed cyclotron orbits driven by $B_z = B\cos\theta$. Hence the total conductivity in this floor interval is 
\be
\sigma^{L,fl} = \sigma^{(1)}_0 \left[1 + \frac{1}{1+(\mu B\cos\theta)^2}\right].
\label{Lfloor}\ee
Although $\rho_{xx}$ is indeed very low, as observed, it is still nearly twice the residual resistivity $\rho_{00} = 1/(2\sigma^{(1)}_0)$.

\subsection{TAMR}
We turn next to the TAMR experiment with $\bf B$ tilted in the plane $k_y$-$k_z$ transverse to $\bf E$ (Fig. \ref{fig4}c). Now, the conversion of states on $C_y$ into looped segments directly suppresses their conductivity. Initially, with $\chi = 0$ (${\bf B}\parallel \bf\hat{z}$), the states $\bf k$ on $C_y$ contribute strongly to $\sigma_{xx}$ despite being parts of hole-type closed orbits. At finite $\chi$, a subset of the planes normal to $\bf B$ intersect $C_y$ to define the surface of a conical wedge (inset in Fig. \ref{fig4}c). As discussed above, the orbits covering the wedge are looped segments that link straight segments on $C_x$ to form open orbits. At the extrema of the loop, the $x$-component of $\bf v(k)$ vanishes. Since ${\bf v}$ appears squared in ${\cal C}_{ab}$ (Eq. \ref{eq:Cab}), this results in a strong suppression of the conductance. In effect, a finite $\chi$ converts high-conductance states on $C_y$ to ones with vanishing conductivity. With increasing $\chi$, the conversion proceeds until it consumes all the high-conduction states on $C_y$. This occurs at the completion angle $\gamma_c\sim 29^\circ$ (Eq. \ref{eq:gammac}). 

Using the tube integral, we have calculated the suppression of $\sigma_{xx}$ in the wedge as a function of $\chi$. For the cylinder $C_y$, the elliptical orbit on the tilted plane can be projected onto a circular orbit ${\cal P}$ in the cross-section of the cylinder (inset in Fig. \ref{fig4}c). On ${\cal P}$, the phase variable $\phi$ then becomes just the azimuthal angle $\varphi$, which greatly simplifies the calculation of ${\cal C}_{ab}$.

As a wavepacket traverses a looped segment, its orbit projects onto an arc of angular length $2\beta$ on $\cal P$. As shown, the angular half-length $\beta_0$ of the longest loop segment is given by
\be
1-\cos\beta_0 = \left(\frac{K}{k_F}-1 \right)\tan\chi.
\ee
We have integrated $0<\beta<\beta_0$ numerically to determine the value of the conductivity $\sigma_{loop}$ at each $\chi$ (Fig. \ref{figA}). The maximum net conductivity from $C_y$ (attained when $\chi = \gamma_c$) is under 0.5$\%$ of that at $\chi = 0$.

Finally, once $\chi$ exceeds $\gamma_c$, the states on $C_y$ abruptly disconnect from open orbits to execute closed cyclotron orbits driven by $B_y = B\sin\chi$, whereas in $C_z$, the closed orbits are driven by $B_z = B\cos\chi$. This holds until $\chi$ increases beyond $\pi/2-\gamma_c$. Then the looped segments wrap around $C_z$ instead of $C_x$, and $\rho_{xx}$ rises steeply. 

The total conductivity in the interval $\gamma_c < \chi < \pi/2 - \gamma_c$ is
\be
\sigma^{T,fl} =  \sigma^{(1)}_0 \left[\frac{1}{1+(\mu B\sin\chi)^2}+ \frac{1}{1+(\mu B\cos\chi)^2}\right].
\label{Tfloor}
\ee
As $\sigma^{T,fl}\ll \sigma^{L,fl}$, the observed resistivity within this interval is
much larger than the floor value in the LAMR scan.

These large-angle features are qualitatively consistent with the experiment. 
A quantitative comparison with $\rho_{xx}$ requires a more involved calculation of $\sigma_{xy}$ (which can be larger than $\sigma_{xx}$).

\section{Sharp resonant features}
To investigate the highly unusual LAMR behavior in the limit of small tilt angles, we have performed high-resolution measurements of $\rho_{xx}$ vs. $\theta$ at fixed $B$. As shown in Fig. 3c, the profile of $\rho_{xx}$ vs. $\theta$ displays a sharp cusp in the limit $\theta\to 0$. This implies that $\rho_{xx}$ deviates from its value at $(0,0)$ in a non-anaylitical way. More interestingly, we observe weak peaks at $\theta = 1.1^\circ$ and $2.2^\circ$. Above the angle $2.2^\circ$, $\rho_{xx}$ steepens its decrease with $\theta$, displaying a sharp break in slope. Because the angular positions of the resonances are independent of $B$, they are unrelated to Landau quantization effects. The tiny $B$-independent angles suggest to us that the features are geometric in origin, arising resonantly at small $\theta$ from very large orbits that extend over multiple Brillouin zones.

A conceptual difficulty in analyzing the small tilt regime is the appearance of quasiperiodic orbits. In Fig. \ref{figB} (Appendix), we plot numerical simulations of the combination of closed and open orbits that appear at small tilt angles $\theta = 1^\circ, 5^\circ$ and $10^\circ$ in the LAMR experiment. In each panel, the plot extends over 25 Brillouin Zones. The orbits are subtly quasiperiodic despite the nominal repetition. As it stands, the tube-intergral approach lacks the formalism to handle quasiperiodic orbit patterns.

\section{Conclusion}
High-resolution angular magnetoresistance performed in the regime $\mu B\gg 1$ in high-mobility metals can uncover novel features that are not evident in conventional Shubnikov de Haas oscillations. In ReO$_3$ with $\mu\sim$ 90,000 cm$^2$/Vs, we observe a singular variation of the resistivity: $\rho_{xx}$ decreases steeply by a factor of 40 when $\bf B$ is tilted in the longitudinal plane containing $\bf E$. However, it rises steeply by a factor of 8-10 when $\bf B$ is tilted in the plane orthogonal to $\bf E$. Using the Shockley tube integral approach, we show that this previously unreported singular variation is inherent to the jungle-gym FS geometry. The AMR profiles display a rounded shoulder at a completion angle $\gamma_c$ that is an intrinsic feature of the FS topology. In addition to explaining $\gamma_c$, the tube-integral approach accounts for the relative magnitudes of the floor values in both the LAMR and TAMR scans.  However, the semiclassical model fails to explain the series of sharp resonant features observed in the LAMR scans (or the cuspy variations as $\theta$ and $\chi$ approach zero). These features, which may involve orbit patterns extending over multiple Brillouin zones, invite further investigation. \\

\centerline{ *      *      *}

\vspace{3mm}
\noindent
{\bf Appendix: Shockley tube integral}\\ \noindent
In general, the semiclassical conductivity in a strong magnetic field $\bf B$ can be computed using the Shockley tube integral~\cite{Ziman,Kagoshima}
\be  
\sigma_{ab} = \frac{2e^2}{(2\pi)^3\hbar^2} \int \frac{m^*}{\omega_c} {\cal C}_{ab}\; dk_H,
\label{sxx}
\ee
where ${\cal C}_{ab}$ is the velocity-velocity correlator discussed below.
The states in $\bf k$ space are divided into a set of parallel planes normal to ${\bf \hat{n}}$ and indexed by $k_H=\bf k\cdot \hat{n}$, where $\bf\hat{n} = B/|B|$. In Eq. \ref{sxx}, $\omega_c$ is the angular frequency of a cyclotron orbit confined to a plane with $m^*$ the cyclotron mass. We may express $m^*$ as the derivative with respect to the energy $\varepsilon$ of the area ${\cal A}$ enclosed by the cyclotron orbit, i.e.
\be
m^* = \frac{\hbar^2}{2\pi} \frac{\partial {\cal A}}{\partial \varepsilon}.
\label{mass}
\ee  
The velocity-velocity correlator ${\cal C}_{ab}$ is given by
\be
{\cal C}_{ab} = \int_0^{2\pi} d\phi \int_0^{\infty}d\phi'
v_a(\phi)v_b(\phi-\phi')\; e^{-\alpha\phi'}.
\ee
Here ${\bf v}(\phi)$ is the group velocity at the phase coordinate $\phi = (\omega_c/eB)\int^{\bf k} dk/v_\perp$ in a cyclotron orbit, with $v_\perp= |{\bf v\times \hat{n}}|$. 

Equation \ref{sxx} is derived using the Green's function of the high-$B$ Boltzmann equation~\cite{Ziman}. The contribution to $\sigma_{ab}$ of a state at the phase coordinate $\phi$ is the sum of wave packets created with velocity $v_b$ by a train of $E$-field $\delta$-function pulses applied at all earlier times corresponding to the phase coordinate $\phi - \phi'$. The wave packets advance along the cyclotron trajectory at the rate $\dot{\phi}' = \omega_c$ while decaying exponentially with the decay constant $\alpha= (\omega_c\tau)^{-1}$ where $\tau$ is the lifetime.

By segmenting the interval $0<\phi'<\infty$ into finite segments, we simplify ${\cal C}_{ab}$ to
\begin{eqnarray}
{\cal C}_{ab} &=& \left(\frac{\hbar k_F}{m_0}\right)^2\frac{1}{(1-e^{-2\pi\alpha})} \times\nonumber\\
  & & \int_0^{2\pi} d\phi \int_0^{2\pi}d\phi' \; v_a(\phi)v_b(\phi-\phi')\; e^{-\alpha\phi'}.
\label{Cab}
\end{eqnarray}

Our goal is to find $\sigma_{xx}$ of the cylinder $C_y$ in a field $\bf B$ tilted at angle $\pi/2-\chi$ to its axis. If we assume the quadratic dispersion $\varepsilon({\bf k}) = \hbar^2(k_x^2+k_y^2)/2m_0$ with band mass $m_0$, Eq. \ref{mass} gives
\be
m^* = m_0/\sin\chi, \quad \alpha = (\omega_c \tau)^{-1} =  (\mu |{\bf B}|\sin\chi)^{-1}.
\label{mass2}
\ee
With $\mu\simeq$ 90,000 cm$^2$/Vs, $\mu B\simeq 81$ at 9 T. 

\begin{figure*}[h]
	\includegraphics[width=0.8\textwidth]{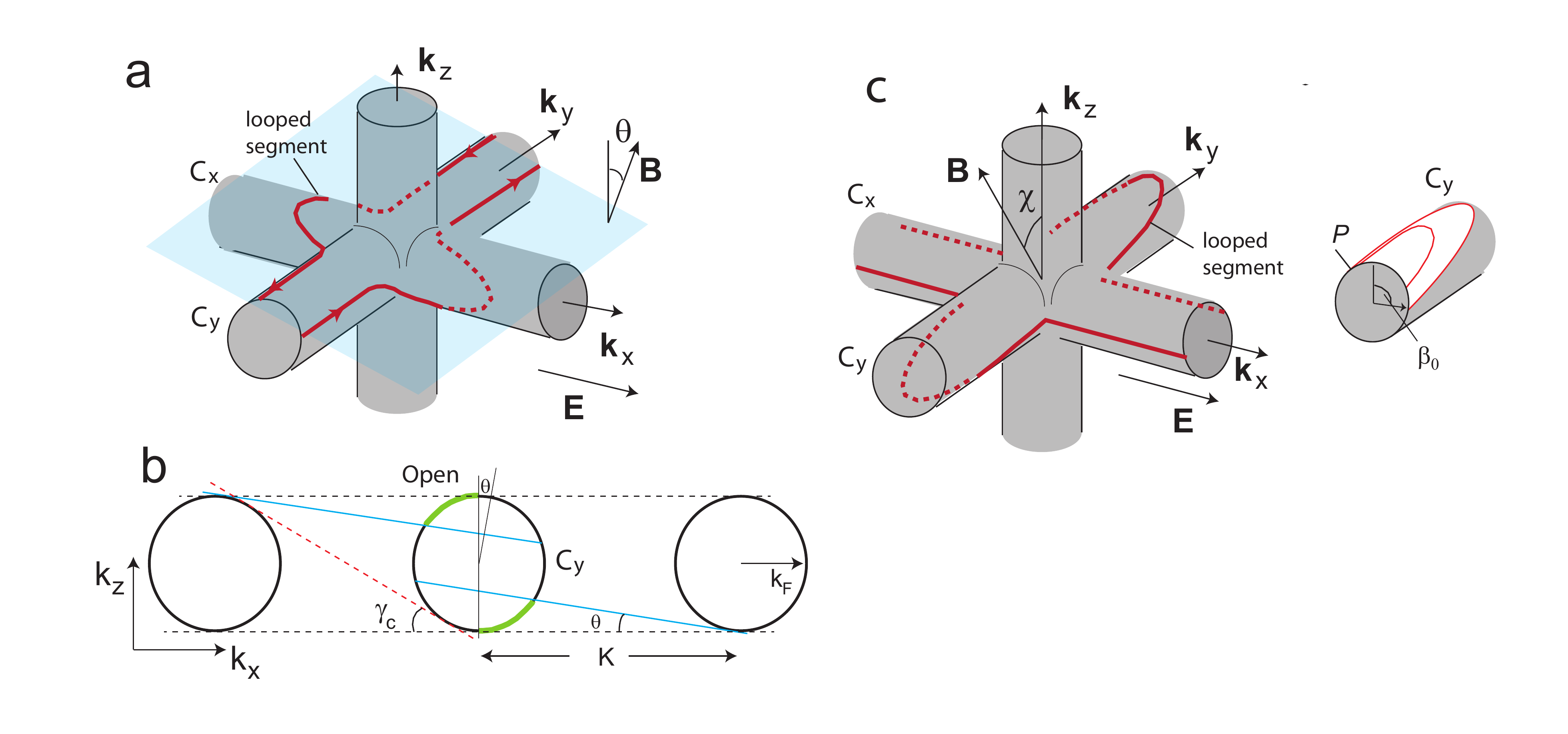}
	\caption{Sketch of open orbits. Panel (a) shows the three FS cylinders $C_x$, $C_y$ and $C_z$ (grey tubes) and a plane normal to $\bf B$ (pale blue). Intersections of the FS with the normal planes define possible orbits of a wave packet. In the LAMR experiment, when $\bf B$ is tilted by $\theta$ relative to $\bf\hat{z}$, an open orbit can emerge (thick red curves). A right-moving wave packet on $C_y$ loops under $C_x$ (dashed curve) before resuming its orbit on $C_y$. The left-moving partner loops over $C_x$. In the high-$B$ limit, these open orbits contribute strongly to $\sigma_{xx}$. Panel (b) shows end-on views of 3 cylinders $C_y$ in the repeated zone scheme with $K= 2\pi/a$. The planes normal to $\bf B$ that are tangential to the outer cylinders (blue lines) define the FS portion hosting open orbits on the middle cylinder (thick green arcs). States outside the green arcs remain in closed orbits. The green arcs lengthen rapidly as $\theta\to \gamma_c$, the completion angle defined by the inner tangent (red dashed line). 
Panel (c): Sketch of open orbits in the TAMR experiment. With $\bf B$ tilted by angle $\chi$ relative to $\bf\hat{z}$ in the plane transverse to $\bf E$, the open orbits are straight-line segments on $C_x$ alternating with looped segments on $C_y$. The inset on the right shows the conical wedge (white area) on $C_y$. Cyclotron orbits on the wedge (red ellipses) project onto circular orbits ${\cal P}$ on the cross-section (front end-face of $C_y$). Each orbit subtends an angle $2\beta$ on ${\cal P}$, while the longest one subtends angle $2\beta_0$. The conductivity arising from states on the entire wedge is obtained by integrating the orbits over the white area (Eq. \ref{Gchi}).
}	
\label{fig4}
\end{figure*}

For the cylinder, the cyclotron period in tilted $\bf B$ is identical to that of a circular orbit ${\cal P}$ projected onto the cross-section in the $k_x$-$k_z$ plane and driven by the field component along $\bf\hat{y}$, $B_y=B\sin\chi$ (inset, Fig. \ref{fig4}c). Moreover, we can replace the phase variable $\phi$ with the azimuthal angle $\varphi$ in ${\cal P}$ (inset in Fig. \ref{fig4}c). At each $\bf k$, $\bf v(k)$ projects to the same vector on ${\cal P}$. 
This simplifies greatly the calculation of $\sigma_{xx}$.\\

\noindent
\emph{Isolated cylinder}\\\noindent
We first consider an isolated cylinder with axis $\parallel \bf\hat{y}$ in a field $\bf B$ tilted at an angle $\chi$ to $\bf\hat{z}$ in the $y$-$z$ plane ($\bf E\parallel\hat{x}$). The cylinder accomodates an electron density 
\be
n_\ell = \frac{2}{(2\pi)^3}\pi k_F^2 K_\ell,
\label{nl}
\ee
where $K_\ell$ is its length. The orbits are closed ellipses with $m^*$ and $\alpha$ given by Eq. \ref{mass2}. 
Integrating $\varphi$ and $\varphi'$ over $(0,2\pi)$ in Eq. \ref{Cab} gives for both ${\cal C}_{xx}$ and ${\cal C}_{zx}$:
\be
{\cal C}_{xx} = \left(\frac{\hbar k_F}{m_0}\right)^2 \frac{\pi\alpha}{1+\alpha^2}, \quad
{\cal C}_{zx} = \left(\frac{\hbar k_F}{m_0}\right)^2 \frac{\pi}{1+\alpha^2} 
\ee
Using these expressions in Eq. \ref{sxx}, the conductivity $\sigma_{xx}$ and the Hall conductivity $\sigma_{xy}$ are
\be
\sigma_{xx} = \frac{n_\ell e\mu}{[1+(\mu B\sin\chi)^2]}, \quad 
\sigma_{zx} = \frac{n_\ell e\mu^2 B\sin\chi}{[1+(\mu B\sin\chi)^2]},
\label{scyl}
\ee
where $\mu = e\tau/m_0$ is the mobility. 

In the limit $\chi\to 0$ ($\bf B\perp$ axis), $\sigma_{xx}$ recovers its zero-$B$ value $n_\ell e\mu$. This is the simplest example of an open-orbit conductivity that is $B$-independent even when $\mu B\gg 1$.\\

\noindent
\emph{Jungle gym FS}\\\noindent
Next, we apply the tube integral to address the TAMR experiment in the jungle-gym FS with intersecting cylinders (Fig. \ref{fig4}c).
Tilting of $\bf B$ in the $k_x$-$k_z$ plane causes a fraction of the hole-like closed orbits to become looped segments that belong to open orbits. The loops are shown as red curves on the white area of the conical wedge in inset of Fig. \ref{fig4}c. In the open orbit, the wave packets traverse alternatingly straight segments on $C_x$ and looped segments on $C_y$ until they damp out. 

As $v_x = 0$ on the former, only the looped segments contribute to $\sigma_{xx}$. Projecting the loop to the circular orbit ${\cal P}$ on the cross-section (inset in Fig. \ref{fig4}c), the azimuthal angle $\varphi$ on ${\cal P}$ runs from $\pi/2-\beta$ to $\pi/2+\beta$ to describe an arc of angular length $2\beta$. Since the planes are indexed by $k_H$, $d\beta$ and $dk_H$ are related by
\be 
dk_H = k_F\cos \chi \sin\beta d\beta.
\label{dkH}
\ee

Evaluating the integrals over $\varphi$ and $\varphi'$ in ${\cal C}_{xx}$ between the limits ($\pi/2-\beta$, $\pi/2+\beta$), we have 
\begin{eqnarray}
{\cal C}_{xx}(\beta) &=& \left(\frac{\hbar k_F}{m_0}\right)^2 
\frac{1}{(1-e^{-2\pi\alpha})} \frac{2e^{-\alpha\pi/2}}{(1+\alpha^2)}   
(\beta - \frac12\sin 2\beta) \times \nonumber\\
&& \left[\alpha\sin\beta\cosh\alpha\beta - 
\cos\beta\sinh\alpha\beta \right].
\end{eqnarray}

As mentioned, the looped segments cover the white area of the conical wedge (inset of Fig. \ref{fig4}c). The longest orbit, corresponding to the maximum angle $\beta_0$, is fixed by the plane tangential to the neighboring $C_y$. Hence $\beta_0$ is determined by
\be
1-\cos\beta_0 = (\Delta K/k_F) \tan\chi,
\label{beta}
\ee
where $\Delta K = K-k_F$. Integrating over all the orbits covering the wedge and using Eq. \ref{n1}, we obtain the conductivity $\sigma^{loop}$ 
\be
\sigma^{loop}(\chi) = n^{(1)} e\mu\frac{k_F}{K-2k_F}\;{\cal G}(\chi),
\label{sigmaloop}
\ee
where ${\cal G}(\chi)$ is the dimensionless integral
\begin{eqnarray}
{\cal G}(\chi) &=& \frac{2}{\pi} \frac{e^{-\alpha\pi/2}}{(1-e^{-2\pi\alpha})}\frac{\alpha\cot\chi }{(1+\alpha^2)} 
\int_0^{\beta_0} (\beta - \frac12\sin 2\beta) \times\nonumber\\
&&		 \left[\alpha\sin\beta\cosh\alpha\beta - 
\cos\beta\sinh\alpha\beta \right]\sin\beta \;d\beta.
\label{Gchi}
\end{eqnarray}\\
${\cal G}(\chi)$ is plotted in Fig. \ref{figA}. As shown, $\sigma^{loop}$ is strongly suppressed. Even when $\chi\to\gamma_c$ (all states on $C_y$ are open orbits), $\sigma^{loop}$ is $< 0.015\times\sigma^{(1)}$. The suppression accounts for the observed increase in $\rho_{xx}$ when $\bf B$ is tilted away from $\bf \hat{z}$ in the TAMR experiment.

\vspace{1cm}
{\bf Acknowledgement}
We have benefitted from discussions with B. A. Bernevig and N. Regnault. RJC and NPO acknowledge support by the U.S. National Science Foundation under award DMR 2011750.

\newpage

%%%%%%%%%%%%%%%%%%%%%
%%%%%%%%%%%%%%%%%%%%%
%%%%%%%%%%%%%%%%%%%%%

\setcounter{figure}{0}
\renewcommand{\thefigure}{S\arabic{figure}}

%%%%%%%%%%%%%%%%%%%%%%%%%%%

\begin{figure}
\includegraphics[width=0.5\textwidth]{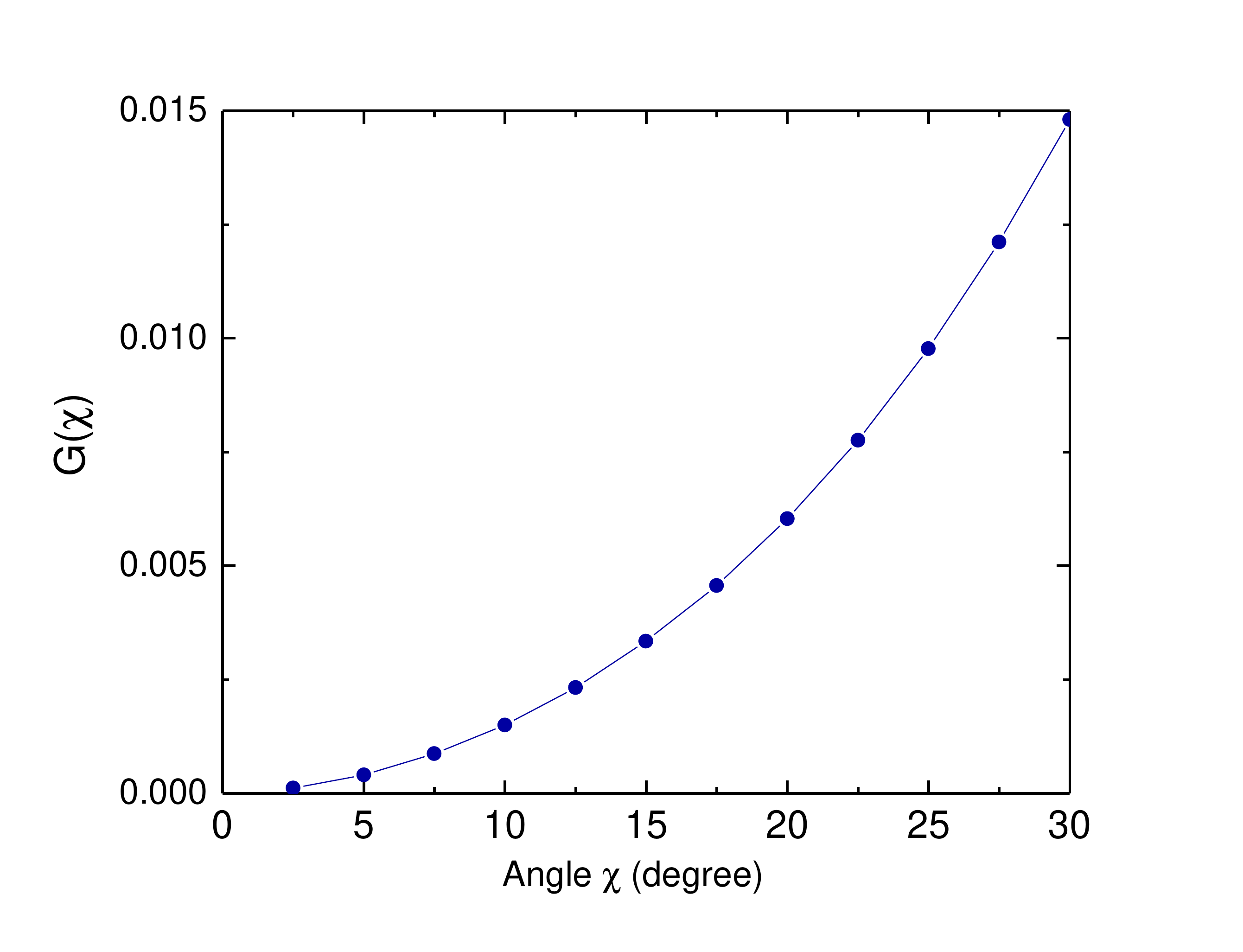}
\caption{\label{figGchi} Variation of the dimensionless integral ${\cal G}$ (Eq. \ref{Gchi}) vs. tilt angle $\chi$. Even when $\chi\to \gamma_c$, ${\cal G}$ is $<$0.015. This implies that the when all the states on $C_y$ are converted to open orbits, its conductivity is suppressed to less than 1.5$\%$ of the value at $\chi$= 0 (see Eq. \ref{sigmaloop}).
}
\label{figA}
\end{figure}

\begin{figure}
\includegraphics[width=0.5\textwidth]{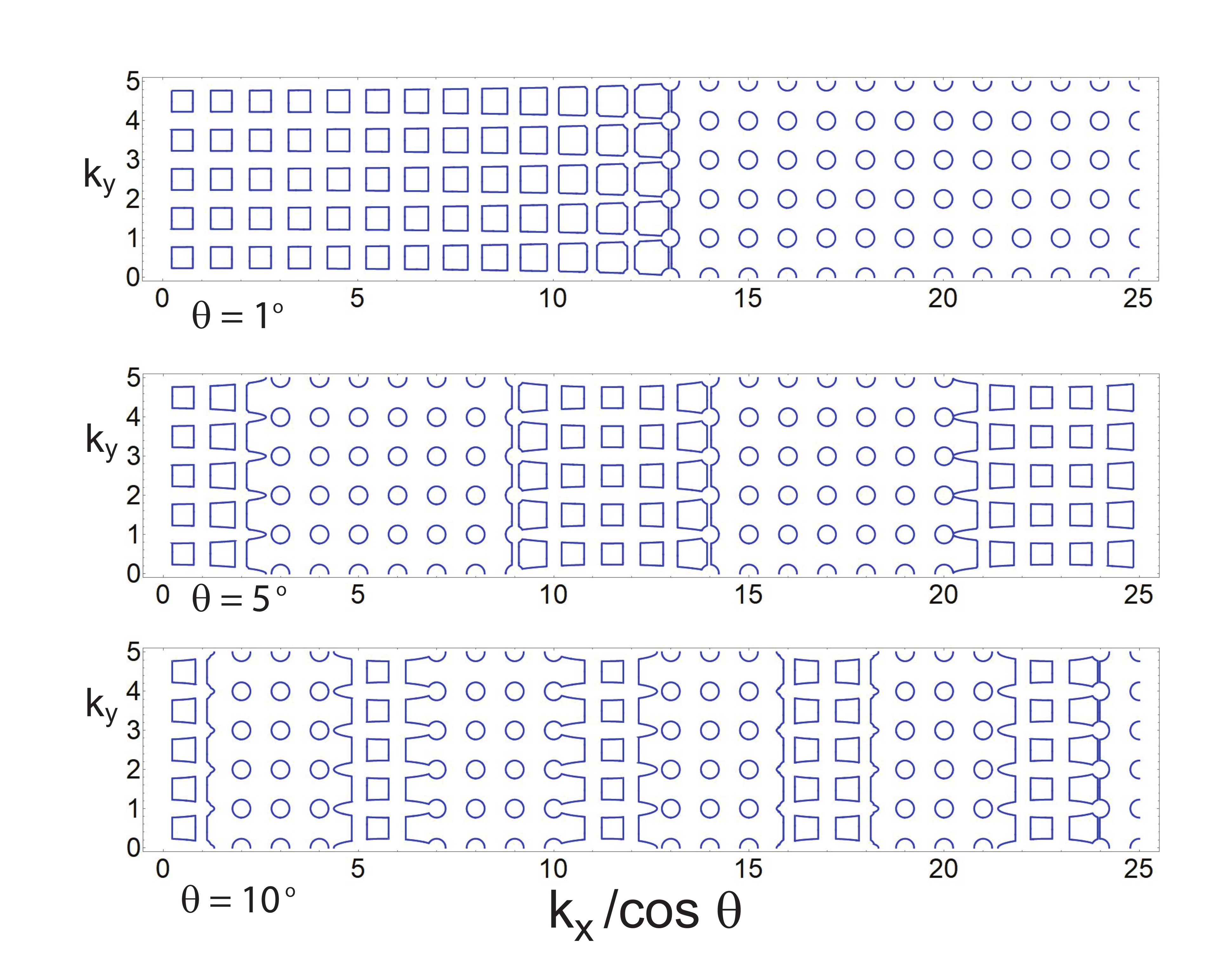}
\caption{\label{figaperiodic} 
Numerical simulation of the pattern of open and closed orbits at three selected values of $\theta$ ($1^\circ, 5^\circ, 10^\circ$) with $\chi = 0$ in the LAMR experiment. The array extends over 25 Brillouin Zones in the extended zone scheme. The orbits lie in a plane normal to $\bf B$, with the horizontal axis $k_x/\cos\theta$ measured in the direction $\bf \hat{z}\times B$. In each panel, the orbits are quasiperiodic despite the appearance of nominal periodicity.
}
\label{figB}
\end{figure}

%%%%%%%%%%%%%%%%%%%%%%%%%%
%%%%%%%%%%%%%%%%%%%%%%%%%%
%%%%%%%%%%%%%%%%%%%%%%%%%%

\end{document}